\documentclass[usenatbib,referee]{mn2e}
\usepackage{epsf}

\newcommand\bF{{\bmath F}}
\newcommand\be{{\bmath e}}
\newcommand\bg{{\bmath g}}
\newcommand\bu{{\bmath u}}
\newcommand\bnabla{{\bmath\nabla}}
\newcommand\p{\partial}
\newcommand\f{\frac}
\newcommand\cst{\mathrm{constant}}
\newcommand\rmd{\mathrm{d}}
\newcommand\rme{\mathrm{e}}
\newcommand\rmi{\mathrm{i}}

\newcommand\real{\mathrm{Re}}
\newcommand\half{{\textstyle{\f{1}{2}}}}
\newcommand\twothirds{{\textstyle{\f{2}{3}}}}
\newcommand\threequarters{{\textstyle{\f{3}{4}}}}

\title[Thermal tides]
{Diurnal Thermal Tides in a Non-synchronized Hot Jupiter}

\author[P.-G. Gu \& G. I. Ogilvie]{Pin-Gao Gu$^{1}$ and Gordon I. Ogilvie$^{2}$\\
$^{1}$Institute of Astronomy \& Astrophysics, Academia Sinica,
Taipei 10617, Taiwan\\
$^{2}$Department of Applied Mathematics \& Theoretical Physics,
University of Cambridge, Wilberforce Road, Cambridge CB3 0WA, UK}

\begin{document}
\maketitle

\label{firstpage}

\begin{abstract}
We perform a linear analysis to investigate the dynamical response
of a non-synchronized hot Jupiter to stellar irradiation. In this
work, we consider the diurnal Fourier harmonic of the stellar
irradiation acting at the top of a radiative layer of a hot
Jupiter with no clouds and winds. In the absence of the Coriolis
force, the diurnal thermal forcing can excite internal waves
propagating into the planet's interior when the thermal forcing
period is longer than the sound crossing time of the planet's
surface. When the Coriolis effect is taken into consideration, the
latitude-dependent stellar heating can excite weak internal waves
(g modes) and/or strong baroclinic Rossby waves (buoyant r modes)
depending on the asynchrony of the planet. When the planet spins
faster than its orbital motion (i.e. retrograde thermal forcing),
these waves carry negative angular momentum and are damped by
radiative loss as they propagate downwards from the upper layer of
the radiative zone. As a result, angular momentum is transferred
from the lower layer of the radiative zone to the upper layer and
generates a vertical shear. We estimate the resulting internal
torques for different rotation periods based on the parameters of
HD 209458b.
\end{abstract}


\section{Introduction}

Hot Jupiters are Jupiter-mass planets located within $\sim 0.1$ AU
from their parent stars. Unlike Jupiter and Saturn in the Solar
System, hot Jupiters are exposed to stellar irradiations that are
much larger than their intrinsic fluxes. Consequently, a deep
radiative outer layer develops on the top of a convective interior
in a hot Jupiter (e.g. see Guillot 2005 for a review).

Infrared observations of hot-Jupiter planetary systems with the
Spitzer Space Telescope have been able to measure temperature
variations and therefore infer temperature distributions on hot
Jupiters \citep{Harrington, Knutson, Cowan}. Meanwhile, a number
of numerical simulations have been developed to investigate
atmospheric circulation on a synchronized or non-synchronized hot
Jupiter to better ascertain the origins of temperature
distributions (see Showman et al. 2007 for a review). Despite the
fact that these simulations are based on different equations and
assumptions, and will thus exhibit different flow features, the
simulated atmospheres usually end up with differential rotations
such as banded structure or vertical shear. Although the flow
patterns deviating from the initial uniform rotation are certainly
the result of planetary rotation, the exact mechanism of how
angular momentum is transported and redistributed between
different regions of the atmosphere is yet to be established.

When a global atmospheric flow follows non-synchronous rotation,
the flow experiences a variation of stellar irradiation which
serves as thermal forcing on the flow, producing thermal tides.
Unlike ocean semi-diurnal tides which are driven by differential
lunar gravity, the semi-diurnal oscillation of the atmospheric
surface pressure\footnote{Diurnal tides of smaller amplitude also
exist in the atmosphere at ground level (Chapman \& Lindzen 1970,
and references therein) but they correspond to a displacement of
the centre of mass of the thermal bulge, which does not contribute
to the gravitational torque (e.g., Correia et al. 2003).} on the
Earth has been known to be mainly excited by the differential
solar heating \citep{Haurwitz}. In a state of quasi-hydrostatic
equilibrium, the gravitational tide in the ocean and solid Earth
and the thermal tide in the atmosphere can be modelled as
gravitational and thermal bulges respectively (see Cartwright 2000
for a historical account). In the case of the Earth, the thermal
bulge and the gravitational bulge have opposite phase difference
with respect to the Sun \citep{Haurwitz,Cartwright}, meaning that
the gravitational torques on the thermal bulge and on the
gravitational bulge are pointing in opposite directions. Since
Venus has a denser atmosphere and receives more solar insolation
than the Earth, thermal tides on Venus are expected to be more
prominent. This idea has inspired a number of models attempting to
explain the slow retrograde spin of Venus by means of a balance
between the torques due to gravitational and thermal tides
\citep{Gold, Dobrovolskis, Correia}. Laskar \& Correia 2004 (cf.
Showman \& Guillot 2002) even postulated that thermal tides may
drive hot Jupiters away from synchronous rotation. This
postulation suggests a mechanism of generating internal tidal heat
in hot Jupiters and may lend support to the tidal inflation model
(Bodenheimer et al. 2001; Mardling 2007 and references therein) in
explaining why some of the transiting hot Jupiters are larger than
indicated by current interior and evolutionary models.

However, thermal bulges are probably not relevant to the case of
gaseous (or liquid) planets. A perfectly rigid crust of a
terrestrial planet can support any atmospheric pressure excess
without being displaced sideways (or being slightly displaced if
the crust is not perfectly rigid; see Corriea \& Laskar 2003). In
the case of gaseous planets, the fluid underlying an overdense
region is freely displaced sideways to attain hydrostatic
equilibrium on the local sound crossing timescale. This means that
any thermally driven density inhomogeneity on the top layer is
almost cancelled out by the density inhomogeneity in the deeper
layers\footnote{One of the easiest ways to understand this concept
is in terms of a planet covered by a liquid ocean and a gaseous
atmosphere.  If a thermal bulge is created in the atmosphere, then
the surface of the ocean is displaced so that the column density
perturbation in the atmosphere at each latitude and longitude is
cancelled by an opposite column density perturbation in the ocean.
In this way, the ocean can remain in hydrostatic equilibrium with
no horizontal pressure gradients, because the same column lies
above every latitude and longitude.}. By this argument, net
thermal bulges cannot form on gaseous planets, and the
gravitational torque acting on the thermal tide is essentially
zero.

Nevertheless, the oscillations of the stellar irradiation can
still excite waves in gaseous planets. It is reminiscent of
dynamical tides in the gravitational tide theories. Waves driven
by gravitational tides in hot Jupiters have been studied in the
literature. Based on the tidal theory by \citet{GN} for high-mass
stars, \citet{Lubow} suggested that the radiative layer of a hot
Jupiter can be tidally synchronized by the internal waves excited
resonantly by the tidal force of the host star. However, in
contrast to high-mass stars where the external irradiation is
unimportant compared to stellar intrinsic luminosity, the stellar
irradiation onto a hot Jupiter is typically several orders of
magnitude stronger than the intrinsic luminosity of the planet. It
implies that the dynamics driven by stellar heating  cannot be
ignored. For instance, internal waves may also be excited
thermally by stellar irradiation on the top of the radiative layer
of a non-synchronized hot Jupiter. In addition, rotation
complicates the behaviour of internal waves. \citet{OL} studied
the internal waves modified by Coriolis forces (i.e. Hough waves)
in hot Jupiters. In the Earth's atmosphere, internal waves of the
diurnal period are restricted in the region of low latitudes where
the Coriolis effect is small, and this explains why the thermal
tide in surface air pressure is predominantly semidiurnal instead
of diurnal \citep{Gold, CL70}. Semi-annual oscillations in
Saturn's low-latitude stratospheric temperatures may be attributed
to wave phenomena driven by seasonal thermal forcing
\citep{Orton}. In the case of a hot Jupiter that is almost tidally
locked by its parent star, the thermal forcing is much slower than
the Coriolis effect, and will likely excite the Rossby waves
(second kind of Hough waves; e.g. see Longuet-Higgins 1968). The
importance of angular momentum transport by internal and Rossby
waves has been discussed in the context of extrasolar giant
planets (see, e.g., Cho 2008). It should be noted that while the
waves driven by gravitational tidal forcing are able to exchange
angular momentum between the planet and its host star, the waves
driven by thermal forcing from the host star on the planet are
only responsible for the angular momentum exchange between
different parts of the planet, because of the cancellation of the
gravitational torque described above.

Atmospheric circulation is an extremely complex topic which
involves turbulence, winds, as well as waves and how they are
thermally driven and interact. Waves driven by thermal tides have
never been studied analytically in the context of hot Jupiters to
understand their basic behaviours. Therefore their roles in
numerical simulations have not been easily identified. In this
paper, we make a first attempt on the wave problem by considering
a ``clean" picture: a diurnal thermal forcing on the radiative
layer with no clouds, winds, turbulence, and gravitational tides.
The radiative flux in the atmosphere is modelled using the
diffusion equation with a power-law Rosseland-mean opacity (cf.
Dobbs-Dixon \& Lin 2008). Although the variation of the stellar
irradiation is not small compared to its mean value, we employ a
linear analysis and investigate the possibility of wave excitation
in a non-synchronized surface layer of a hot Jupiter driven by
stellar irradiation. The goal is to estimate how much angular
momentum can be redistributed by thermal tides near the surface of
a hot Jupiter in our simple linear theory. We first focus on the
thermal tide problem for internal waves in a non-rotating
plane-parallel atmosphere in \S2. Then we turn our study to Hough
waves in a rotating atmosphere in the form of a spherical shell in
\S3. Finally, the results are summarized and discussed in \S4.

\section{The non-rotating plane-parallel atmosphere}

\subsection{Basic equations}

We initially consider a non-rotating plane-parallel atmosphere
with uniform gravity $\bg=-g\,\be_z$.  The fluid equations for an
ideal gas are
\begin{equation}
  \f{\p\bu}{\p t}+\bu\cdot\bnabla\bu=-\f{1}{\rho}\bnabla p+\bg,
\end{equation}
\begin{equation}
  \f{\p\rho}{\p t}+\bnabla\cdot(\rho\bu)=0,
\end{equation}
\begin{equation}
  \f{\p p}{\p t}+\bu\cdot\bnabla p+\gamma p\bnabla\cdot\bu=
  -(\gamma-1)\bnabla\cdot\bF,
\end{equation}
\begin{equation}
  \bF=-\f{16\sigma T^3}{3\kappa\rho}\bnabla T,
\label{f}
\end{equation}
\begin{equation}
  p=\f{R\rho T}{\mu},
\end{equation}
where $\bu$ is the fluid velocity, $p$ is the gas pressure, $\rho$
is the mass density, $T$ is the temperature, $\bF$ is the
radiative flux density, $\kappa$ is the opacity, $\mu$ is the mean
molecular weight, $R$ is the gas constant, $\sigma$ is the
Stefan-Boltzmann constant, and $\gamma$ is the ratio of specific
heats. For simplicity we assume that $\gamma$ and $\mu$ are
constant. We use the radiative diffusion approximation~(\ref{f})
throughout the atmosphere and apply the `Marshak' boundary
condition (cf.~Pomraning 1973)
\begin{equation}
  \sigma T^4=\half F_z+F_\rmi
  \label{eq:Marshak}
\end{equation}
at $z=+\infty$, where $F_\rmi$ is the irradiating flux. The
extension of the radiative diffusion approximation to the
optically thin atmosphere is done for the sake of simplicity and
is clearly a limitation of our model.

\subsection{Equilibrium state with a power-law opacity}

We consider an equilibrium reference state consisting of a static
atmosphere that is uniformly irradiated by the mean stellar
irradiation. For the equilibrium state we have
\begin{equation}
  \f{\rmd p}{\rmd z}=-\rho g,
\end{equation}
\begin{equation}
  F_z=-\f{16\sigma T^3}{3\kappa\rho}\f{\rmd T}{\rmd z}=\cst,
\end{equation}
where $F_z$ is the intrinsic radiative flux density of the planet.
Let $\tau$ be the optical depth measured from $z=+\infty$.  Then
$\rmd\tau=-\kappa\rho\,\rmd z$ and we have
\begin{equation}
  \f{\rmd p}{\rmd\tau}=\f{g}{\kappa},
\label{dpdtau}
\end{equation}
\begin{equation}
  \f{\rmd}{\rmd\tau}(\sigma T^4)=\threequarters F_z.
\label{dtdtau}
\end{equation}
The solution of eq.~(\ref{dtdtau}) subject to the boundary
condition (\ref{eq:Marshak}) is
\begin{equation}
  \sigma T^4=\threequarters F_z(\tau+\twothirds)+F_\rmi.
\end{equation}
Note that $\sigma T^4=F_z+F_\rmi$ at the photosphere
$\tau=\twothirds$.

The equation for hydrostatic equilibrium can be analytically
solved if we assume a power-law opacity:
\begin{equation}
\kappa=c_\kappa p^a T^{-4b}
\end{equation}
for constants $a$, $b$, and $c_\kappa$.  Then
\begin{equation}
  \f{p^a}{T^{4b}}\f{\rmd p}{\rmd T^4}=\f{4\sigma g}{3c_\kappa F_z}.
\end{equation}
The solution satisfying $p=0$ at $\tau=0$ (where $T=T_\infty$) is
\begin{equation}
  \f{p^{a+1}}{a+1}=\f{4\sigma g}{3c_\kappa F_z}\left(\f{1}{b+1}\right)\left[T^{4(b+1)}-T_\infty^{4(b+1)}\right].
\label{eq:pT}
\end{equation}

The top of the convective layer is located where the Schwarzschild
criterion for marginal stability is satisfied; i.e., setting the
Brunt--V\"ais\"al\"a frequency $N=g^{1/2}[(1/\gamma)\mathrm{d} \ln
p/\mathrm{d} z -\mathrm{d} \ln \rho /
  \mathrm{d} z]^{1/2}$ equal to zero gives
\begin{equation}
  \f{\rmd\ln p}{\rmd\ln T}=\f{\gamma}{\gamma-1}
\end{equation}
at $\tau=\tau_{conv}$.
Thus
\begin{equation}
  \left(\f{T_{conv}}{T_\infty}\right)^{4(b+1)}=\f{\gamma (a+1)}{\gamma(a+1)-4(b+1)(\gamma-1)}
  \equiv X^{b+1}.
\end{equation}
We require the denominator to be positive for convection to start.
Since
\begin{equation}
  \f{T^4}{T_\infty^4}=\f{\threequarters F_z(\tau+\twothirds)+F_i}{\half F_z+F_i},
\end{equation}
we obtain
\begin{equation}
  \tau_{conv}=\f{4(F_z/2+F_i)}{3F_z}(X-1).
\label{eq:tau_conv}
\end{equation}
If we treat $\kappa$ as a constant ($a=b=0$), convection does not
occur for $\gamma \geq 4/3$. In this paper, the linear analysis
will be performed for the radiative layer sandwiched by the top
boundary at $\tau=0$ and the bottom boundary at
$\tau=\tau_{conv}$.

Having found $T(\tau)$ and $p(\tau)$, we have $\rho(\tau)$ and can
then solve for $z(\tau)$.  However it is more convenient just to
use $\tau$ instead of $z$ as a vertical coordinate in the problem.
The solution is completely determined once the parameters $g$,
$c_\kappa$, a, b, $\mu$, $F_z$ and $F_\rmi$ are specified.

\subsection{Linear perturbation analysis}

We consider Eulerian perturbations of the form
\begin{equation}
  \real\left[\bu'(z)\,\rme^{\rmi k_x x-\rmi\omega t}\right],
\end{equation}
etc., where $k_x$ is a real horizontal wavenumber, $x$ is the
horizontal Cartesian coordinate, and $\omega$ is a real frequency
of the thermal forcing.
In this paper, we shall consider a hot Jupiter in a circular orbit
with the orbital period $2\pi/n_{orb}$ and consider that its spin
axis is normal to the orbital plane, although in this section we
neglect the dynamical effects of rotation. The thermal tide is
driven by a variation of the irradiating flux, and the problem at
hand is to work out the amplitude and phase of the perturbations
that result.

The linearized equations read
\begin{equation}
  -\rmi\omega u_x'=-\f{\rmi k_xp'}{\rho},
\end{equation}
\begin{equation}
  -\rmi\omega u_z'=-\f{1}{\rho}\p_z p'+\f{\rho'}{\rho^2}\p_z p,
\end{equation}
\begin{equation}
  -\rmi\omega\rho'+u_z'\p_z\rho+\rho(\rmi k_xu_x'+\p_zu_z')=0,
\end{equation}
\begin{equation}
  -\rmi\omega p'+u_z'\p_z p+\gamma p(\rmi k_xu_x'+\p_zu_z')=
  -(\gamma-1)(\rmi k_xF_x'+\p_zF_z'),
\end{equation}
\begin{equation}
  F_x'=F_z\left(\f{\rmi k_xT'}{\p_zT}\right),
\end{equation}
\begin{equation}
  F_z'=F_z\left(\f{\p_zT'}{\p_zT}+\f{3T'}{T}-\f{\rho'}{\rho}-\f{\kappa'}{\kappa}\right),
\end{equation}
\begin{equation}
  \f{p'}{p}=\f{\rho'}{\rho}+\f{T'}{T}.
\end{equation}
This system of ODEs is of fourth order and the dependent variables
can be taken as $\xi_z$, $p'$, $T'$ and $F_z'$, where $\xi_z$ is
the vertical displacement given by $u_z'=-\rmi\omega\xi_z$.
Rewriting $\p_z=-\kappa\rho\,\p_\tau$, we obtain the system
\begin{equation}
  \p_\tau\xi_z=(\p_\tau\ln T-\p_\tau\ln p)\xi_z-
  \f{k_x^2p'}{\omega^2\kappa\rho^2}+
  \f{1}{\kappa\rho}\left(\f{p'}{p}-\f{T'}{T}\right),
  \label{l1}
\end{equation}
\begin{equation}
  \p_\tau p'=-\f{\omega^2\xi_z}{\kappa}+
  \f{g}{\kappa}\left(\f{p'}{p}-\f{T'}{T}\right),
  \label{l2}
\end{equation}
\begin{equation}
  \p_\tau T'=\left[(a+1)\f{p'}{p}-(b+1)\f{4T'}{T}+\f{F_z'}{F_z}\right]\p_\tau T,
  \label{l3}
\end{equation}
\begin{equation}
  \p_\tau F_z'=p\left[\left(\f{\gamma}{\gamma-1}\right)\p_\tau\ln T-
  \p_\tau\ln p\right]\rmi\omega\xi_z+
  \f{k_x^2F_zT'}{\kappa^2\rho^2\p_\tau T}+
  \f{\rmi\omega p}{\kappa\rho}\left[\f{p'}{p}-
  \left(\f{\gamma}{\gamma-1}\right)\f{T'}{T}\right].
  \label{l4}
\end{equation}
In the Appendix, we argue, using a scale analysis and a
dimensional reduction of the problem, that the first term on the
right hand side of eq.~(\ref{l2}) and the second term on the right
hand side of eq.~(\ref{l4}) can be neglected. Neglecting these
small terms amounts to assuming vertical hydrostatic balance and
neglecting horizontal radiative diffusion. The large scales are
also neglected since the geometry is planar and there is no
rotation.

The above four ODEs can be solved once four boundary conditions
are given. In our model, we assume a thermal balance among
perturbed energy fluxes at the top boundary; i.e., linearizing the
Marshak boundary condition eq.~(\ref{eq:Marshak}) gives
\begin{equation}
4\sigma T^3 T'= \f{1}{2} F'_z + F'_i \label{eq:pert_flux}
\end{equation}
at $\tau=0$. In other words, the thermal forcing, which is the
perturbed irradiation $F'_i$, is introduced to the system via the
top boundary conditions. In the Appendix, we describe the
mathematical details of how we determine the second boundary
condition associated with the singular point at $\tau=0$.

To specify $F'_i$,
we assume that as the planet rotates, the stellar irradiation changes sinusoidally during the
day and is completely switched off during the night.
In the plane-parallel case, the stellar irradiation (heating term) is
then proportional to
\begin{equation}
  \cos\tilde\phi\,H(\cos\tilde\phi).
\end{equation}
where $H$ is the Heaviside step function and $\tilde\phi$ is the
longitude measured in a frame rotating with the orbit relative to
the substellar point; namely,
$\tilde\phi=x/R_p-(n_{orb}-\Omega)t$. The above thermal variation
can be decomposed into a Fourier series in $\tilde\phi$ as
follows:
\begin{equation}
  \cos\tilde\phi\,H(\cos\tilde\phi)=\f{1}{\pi}+\f{1}{2}\cos\tilde\phi+
  \f{2}{3\pi}\cos2\tilde\phi+\{m=4\hbox{ terms and above}\},
  \label{eq:thermalforcing_plane}
\end{equation}
where $m$ is the azimuthal wavenumber. The first term (i.e. $m=0$)
of the Fourier components is steady. It produces no tide but
provides the uniform irradiating flux $F_i$. Other terms in the
above equation give rise to the perturbed oscillatory irradiation
$F'_i$. In this paper, we only consider the diurnal oscillatory
component (i.e. $m=1$, the second term on the right hand side of
eq.~(\ref{eq:thermalforcing_plane})) for $F'_i$. In other words,
the amplitude of $F'_i$ is $(\pi/2)F_i$, $\omega=n_{orb}-\Omega$,
and $k_x=1/R_p$ for a planet of radius $R_p$ and spin rate
$\Omega$.

The other two boundary conditions at $\tau = \tau_{conv}$ are
dependent on how the dynamics of the atmospheric gas varies with
the thermal forcing frequency $\omega$ (see the explanations
following eq.~(\ref{eq:pgmode}) for more details). Let $c_{sp}$ be
the isothermal sound speed at the photosphere. Then the inverse of
the horizontal sound-crossing time of the photosphere between the
day and night sides of the planet is $c_{sp}/\pi R_p$. In the case
of diurnal forcing, when $|\omega| \gg c_{sp}/\pi R_p$, the
solutions of the ODEs behave like thermal diffusion and are
expected to decay quickly with depth. On the other hand, when
$|\omega| \ll c_{sp}/\pi R_p$, the equations admit a solution in
the form of an internal wave which propagates downwards. If the
depth of the radiative layer is large enough, the internal waves
can decay quickly due to radiative loss before the wave reaches
the turning point where $N=|\omega|$. Therefore in these two
dynamical limits, we can set
\begin{equation}
p'=T'=0 \label{eq:bbc1}
\end{equation}
at $\tau=\tau_{conv}$. We shall see later in this paper that the
turning point of internal waves is extremely close to the bottom
of the radiative zone. Hence when $|\omega| \approx c_{sp}/\pi
R_p$ and thereby enabling the internal waves to propagate to the
turning point, these waves are not expected to have completely
decayed at $\tau_{conv}$. The boundary conditions $p'=T'=0$ may
not be appropriate at $\tau_{conv}$ in this dynamical regime, and
we should properly continue the wave solution into the convective
region below. In this paper, we restrict ourselves primarily to
the applications for large and small $\omega$ in the
plane-parallel case. We note that setting $p'=0$ at the bottom
boundary imposes the condition that the perturbed column density
above the bottom boundary is zero as a result of the vertical
hydrostatic equilibrium; i.e., this eliminates any thermal bulges
in our calculations. As we have already explained in the
Introduction, without a hard surface thermal bulges are unlikely
to form on a gaseous planet.

In summary, in the plane-parallel case we aim to solve the 4
linearized ODEs (\ref{l1})-(\ref{l4}). At $\tau \ll 1$, we apply
the boundary conditions associated with the imposed thermal
forcing eq.~(\ref{eq:pert_flux}) (see
eq.~(\ref{eq:tbc1})-eq.~(\ref{eq:tbc4}) in the Appendix for
details). At $\tau =\tau_{conv}$, we adopt the boundary conditions
eq.~(\ref{eq:bbc1}) which are valid for $|\omega| \gg$ or $\ll
c_{sp}/\pi R_p$.

\subsection{Numerical Results}

We consider an atmosphere with no clouds for simplicity in this
paper, We choose HD 209458b as an illustrative example for the
thermal-tide study in the paper because the intrinsic luminosity
$F_z=3.76\times 10^6$ erg/cm$^2$ can be obtained from the
simulation for the interior structure of HD 209458b in the
grain-free case (Bodenheimer, private communication). Some
internal heating has been applied to the interior-structure model
to explain the observed radius $R_p=1.32R_J$ of HD 209458b
(Bodenheimer et al. 2003). We also employ the following input
parameters for HD 209458b in solving the linearized equations:
$M_p=0.69M_J$,
$F_i=2.22\times 10^8$ erg/cm$^2$ s, $2\pi/n_{orb}=3.52474859$
days, $\mu=2$ g/mol., and $\gamma=1.4$ for the gas in the
radiative surface layer of the planet.

\begin{figure}
\epsfxsize=17 cm \epsffile{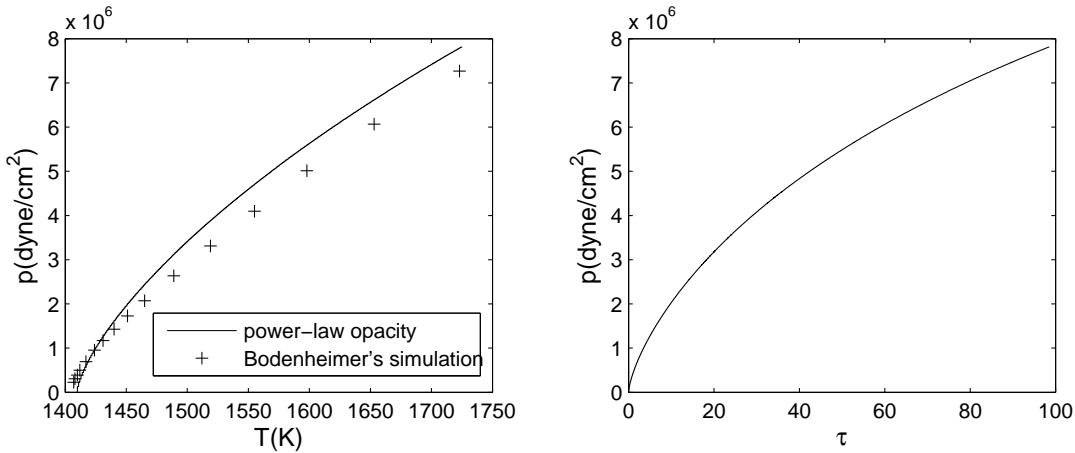} \caption{The structure of the
uniformly irradiated equilibrium reference state for the radiative
layer of HD 209458b. In the left panel, the data from
Bodenheimer's simulation (cross points) based on the opacity table
by Freedman et al. 2008 are compared to the analytical solution of
eq.~(\ref{eq:pT}) using the power-law opacity eq.~(\ref{eq:kappa})
(solid curve). In the right panel, the relation between the
pressure and the optical depth $\tau$ derived from the analytical
solution is shown.} \label{fig:powerlaw}
\end{figure}

Without grains, we are able to
fit the molecular Rosseland opacity computed
by Freedman et al. (2008)
suitable for the radiative layer of a hot Jupiter (i.e. $p\approx
1$ bar and $T\approx 1600$K) to the power-law:
\begin{equation}
\kappa=8.39 \times 10^{-3} \left( {p \over 1 {\rm bar}} \right)^a
\left( {T\over 1600 {\rm K}} \right)^{-4b}\ {\rm cm^2/g} =
c_\kappa p^a T^{-4b}, \label{eq:kappa}\end{equation} where
$a=0.42$ and $b=-0.987$. $c_\kappa=10^{-17.27}$ when expressed in
CGS units. The fidelity of applying this power-law opacity can be
justified by comparing with Bodenheimer's grain-free simulation.
Figure~\ref{fig:powerlaw} illustrates the comparison: the left
panel shows that the structure of the radiative layer for HD
209458b from Bodenheimer's simulation agrees closely with the
equilibrium state described by eq.~(\ref{eq:pT}) using the
power-law opacity. We also apply this power-law opacity to the
optically thin atmosphere $\tau<2/3$. This is done for the sake of
simplicity but is certainly a limitation of our model. We note
that the radiative layer in the grain-free model is shallower than
that in other interior-structure models (e.g., see Guillot 2005).
It is because $R_p$ and $M_p$ being the same, $F_z$ increases as
the opacity in the radiative layer decreases, resulting in a
thinner radiative layer according to eq.~(\ref{eq:tau_conv}). In
the following, the vertical structure of solutions will be
presented as a function of $\tau$. However, the readers can refer
to the right panel of Figure~\ref{fig:powerlaw} to convert the
$\tau$ coordinate to the pressure coordinate, for the case of HD
209458b.

We set the positive $x$-direction as the direction of the planet's
rotation. We focus on the case that the planet is rotating faster
than its orbital motion. Therefore the thermal forcing and the
thermal tides propagate in a retrograde sense; i.e., the
substellar point moves backwards in the frame of the rotating
planet ($\omega<0$).

\begin{figure}
\epsfxsize=17 cm \epsffile{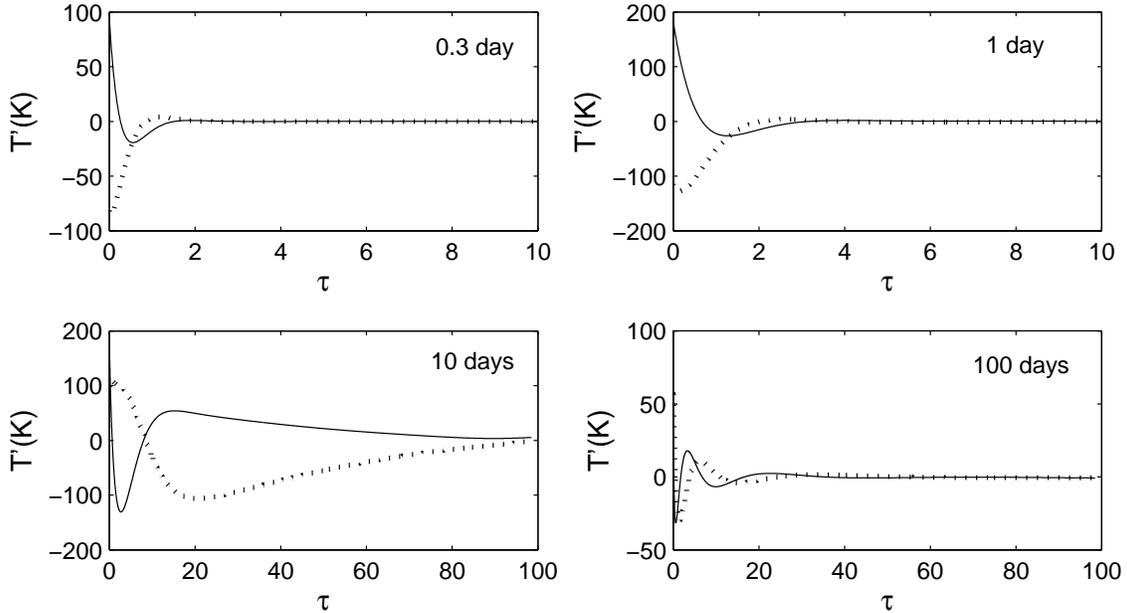} \caption{Temperature
perturbation $T'$ as a function of $\tau$ for the cases of 4
different forcing periods: $-2\pi/\omega=$ 0.3, 1, 10, and 100
days. The real and imaginary parts of $T'$ are denoted by a solid
and a dotted curve respectively.} \label{fig:T}
\end{figure}

\begin{figure}
\epsfysize= 7 cm \epsfxsize=17 cm \epsffile{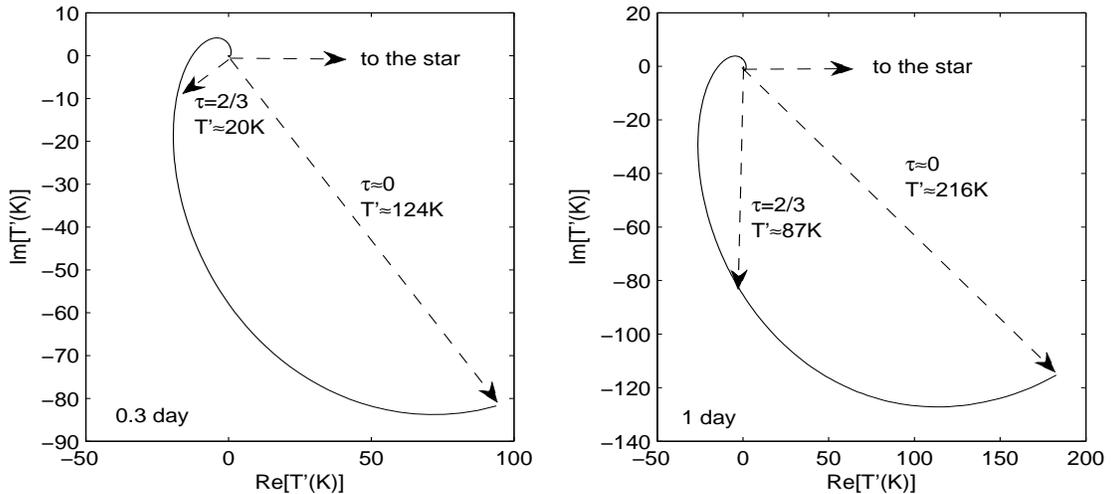}
\caption{Phase diagrams of $T'$ in the complex-number plane for
the cases of two forcing periods: 0.3 day (left panel) and 1 day
(right panel). The phase vector $T'$ spirals clockwise from
$\tau\approx 0$ to large $\tau$, meaning that $T'$ exhibits a
larger phase lag with respect to the star (i.e. the direction of
the thermal forcing as shown pointing to the right in the
horizontal direction) as $\tau$ increases, its magnitude
decreasing and finally dropping to zero. The phase vector $T'$ for
$\tau\approx 0$ (i.e. at the top of the atmosphere) and $\tau=2/3$
(i.e. at the photosphere) are shown.} \label{fig:phase}
\end{figure}

Figure~\ref{fig:T} shows the vertical structure of $T'$ in units
of K for different thermal forcing periods $2\pi/\omega$ ranging
from 0.3 day (a case for a fast rotating planet) to 100 days (a
case close to the synchronous state). The real and imaginary parts
of $T'$ are denoted by a solid and a dotted curve respectively.
Although the results show that $|T'| \ll T$ in all cases,
$|dT'/dz|$ is not smaller than $|dT/dz|$ especially at small
$\tau$ owing to the nonlinear forcing (i.e. $|F'_i| \sim F_i$ at
the top boundary). Therefore our linear analysis is less
justified.

When the forcing periods are short (e.g. 0.3 and 1 day as shown in
the top two panels), $T'$ decays with depth and the solutions
behave like those to the thermal diffusion problem with the heat
diffusing from the top of the atmosphere to a depth characterized
by the diffusion length $\approx \sqrt{ 2D /|\omega |}$, where $D$
is the thermal diffusion coefficient of the atmosphere. Comparing
the 0.3-day to the 1-day case, Figure~\ref{fig:T} shows that $T'$
can penetrate deeper in the 1-day case as a result of a longer
forcing period and therefore a longer diffusion length. The
phenomenon of thermal diffusion can be also verified by the phase
diagram of the complex number $T'$. Figure~\ref{fig:phase} shows
that in the cases of the forcing periods $=0.3$ and 1 day, the
solutions for $T'$, denoted by the solid curve in the real complex
plane, spiral clockwise toward the origin as $\tau$ increases. The
direction of the thermal forcing is shown by the horizontal arrow
pointing to the right. For the forcing $\propto \exp({-i\omega
t})$ with $\omega <0$, this means that the peak value of the
perturbed temperature $|T'|$ exhibits a phase lag (i.e. delay)
with respect to the star. Furthermore, while the phase lag
increases with depth, $|T'|$ decreases with depth. All of these
results demonstrate the process of thermal diffusion.

\begin{figure}
\epsfxsize=13cm \epsfysize=7cm \epsffile{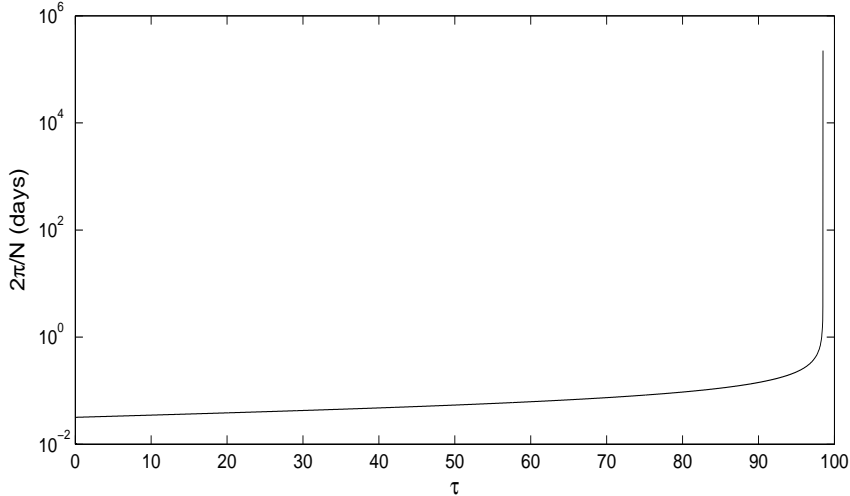} \caption{The
Brunt--V\"ais\"al\"a frequency as a function of $\tau$ resulting
from the vertical structure with the use of the power-law opacity
shown in Figure~\ref{fig:powerlaw}. In this case, convection
occurs when $\tau_{conv}\approx 98.5$.} \label{fig:Brunt}
\end{figure}

On the other hand, when the forcing periods are long (e.g. 10 and
100 days as shown in the bottom panels of Figure~\ref{fig:T}), the
vertical profiles of $T'$ exhibit wavelike solutions, meaning that
waves are excited from the top of the atmosphere and propagate in.
These waves are known as internal waves (i.e. g modes). The
dispersion relation of g-mode oscillation in the WKB linear
perturbation analysis without dissipation reads
\begin{equation}
k_x^2+k_z^2 = {N^2 \over \omega^2} k_x^2, \label{eq:dispersion}
\end{equation}
where $k_x$ and $k_z$ are the horizontal and the vertical wave
numbers respectively. Figure~\ref{fig:Brunt} shows the vertical
profile of $N$ for our parameters for the radiative layer of HD
209458b. $N\gg |\omega|$ throughout the radiative layer except for
the region very close to $\tau=\tau_{conv}\approx 98.5$. The
dispersion relation indicates that for a given $k_x$, the vertical
wavelength $2\pi /k_z$ decreases with the forcing period $2\pi/
\omega$. Furthermore, internal waves can be dissipated due to
radiative loss especially in the top layer of the radiative zone
where the thermal timescale is short and $k_z$ is large. The
shorter the wavelength is, the faster the wave is radiatively
dissipated. This is exactly what is shown in the bottom panels of
Figure~\ref{fig:T}. The internal wave for the 100-day case has a
much shorter wavelength and hence decays faster with depth than
the wave for the 10-day case.

The different dynamics appearing in the cases of short (thermal
diffusion) and long (g mode) forcing periods can be also
understood by comparing the forcing period with the sound crossing
time of the planet's surface.
When the forcing period is
longer than the sound crossing time of the planet's surface; i.e.,
\begin{equation}
{2\pi \over |\omega|} > {\pi \over  k_x c_{sp}} = {\pi R_p \over
c_{sp}} \approx 1.5 \left( {R_p \over 1.32 R_j} \right) \left(
{T\over 1500{\rm K}} \right)^{-1/2} \left( {\mu \over 2}
\right)^{-1/2}\  {\rm days}, \label{eq:pgmode}
\end{equation}
downward-travelling internal waves (incompressible modes) can be
driven by the diurnal forcing. On the other hand, when $|\omega| >
k_x c_{sp}/2$, the day-side and the night-side are causally
disconnected\footnote{Although the day and the night sides are
causally disconnected, the vertical hydrostatic balance of
perturbations is still valid (see the Appendix).}. The diurnal
forcing results only in the thermal diffusive effect in the
vertical direction.


\begin{figure}
\epsfxsize 17 cm \epsffile{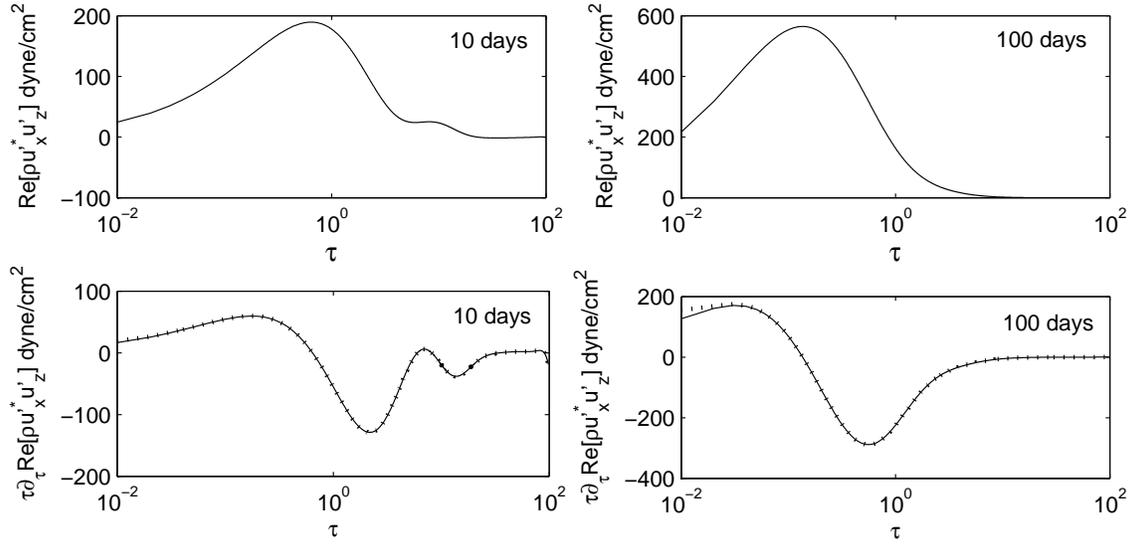} \caption{The top two
panels show the momentum flux Re[$\rho {u'_x}^* u_z'$] as a
function of $\tau$ for the cases of two forcing periods: 10 (left
panel) and 100 (right panel) days. In the bottom two panels, the
gradient of the momentum flux multiplied by $\tau$, denoted by the
dotted curve, is plotted for comparison with the solid curve
described by the right-hand side of eq.~(\ref{eq:radidamp}).}
\label{fig:momentum}
\end{figure}

In the plane-parallel case, internal waves carry a vertical flux of
horizontal momentum. This flux builds up in the upper layers of the
atmosphere where the waves are thermally forced, and is reduced at
greater depth where the waves are radiatively damped and transfer
their momentum to the atmosphere. It
can be shown from the linearized equations (\ref{l1})-(\ref{l4})
that the gradient of the vertical momentum flux density is related to the
non-adiabatic term $\nabla \cdot \bF'$ as follows:
\begin{equation}
\tau \p_\tau \left[ {\rm Re} ( \rho u'_x {u'_z}^*) \right] =
-\f{\tau k_x}{\kappa \rho \omega}{\rm Re}\left[
\f{(\gamma-1)}{\rmi \omega} \left( \nabla
 \cdot \bu' \right)^* \left(
\nabla \cdot \bF' \right)\right]. \label{eq:radidamp}
\end{equation}
 Figure~\ref{fig:momentum} shows the
vertical profiles of the momentum flux Re[$\rho {u'_x}^* u_z'$]
and its gradient for the cases of two forcing periods: 10 and 100
days. The bottom two panels indicate that the gradient of the
momentum flux, which has been multiplied by $\tau$ for clearer
illustration, agrees with the right-hand side of
eq.~(\ref{eq:radidamp}). This validates the relation between
momentum transfer and radiative damping in our result. General
speaking, both cases indicate that the momentum flux is positive
and indicate that the momentum is transported from the inner
region where the gradient of the momentum flux is negative to the
outer region where the gradient of the momentum flux is positive.
In other words, the downward-travelling internal waves excited by
the thermal forcing transport momentum outward. In the 10-day
case, however, the situation is more complicated since the
internal waves can penetrate deeper as a result of less damping
due to longer vertical wavelength (see Figure~\ref{fig:T}).

\begin{figure}
\epsfxsize 17 cm \epsfysize 7.6 cm \epsffile{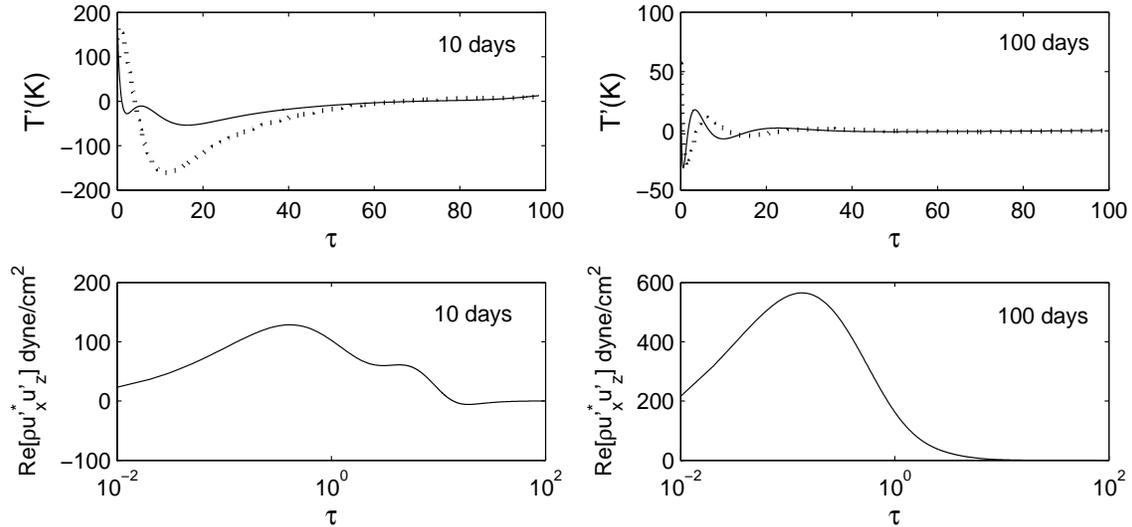}
\caption{Vertical profiles of $T'$ (the solid and the dotted curve
denote the real and imaginary part respectively) and the vertical
momentum flux Re[$\rho {u'_x}^* u_z'$] for $2\pi/\omega$=10 and
100 days based on the following bottom conditions at
$\tau_{conv}$: $p'/p=\gamma \rho'/\rho$ and the Lagrangian density
perturbation =0.} \label{fig:bc2}
\end{figure}

We note that the undamped internal waves in the 10-day case reach
the turning point where $N=\omega$ (and therefore $k_z=0$) near
the radiative-convective boundary (see eq.~(\ref{eq:dispersion})).
The waves are evanescent beyond the turning point and are
reflected so as to interfere with the downward-travelling waves.
As is clear from Figure~\ref{fig:Brunt}, the turning point is
extremely close to the bottom of the radiative layer, implying
that internal waves may not be evanescent significantly at
$\tau_{conv}$ where we nevertheless have imposed the boundary
conditions $p'=T'=0$. The consequence is that the solution in the
10-day calculation is sensitive to the bottom boundary conditions
at $\tau_{conv}$. To demonstrate this point, we consider the
possibility that the perturbations at $\tau_{conv}$ may still
preserve both the adiabatic and incompressible properties of
internal waves instead of the ``decaying" conditions $p'=T'=0$. We
apply alternative bottom boundary conditions to the 10- and
100-day cases: $p'/p=\gamma \rho'/\rho$ (adiabatic perturbation)
and the Lagrangian density perturbation =0 (incompressible
perturbation). The results are shown in Figure~\ref{fig:bc2}.
Comparing with Figure~\ref{fig:T} and Figure~\ref{fig:momentum},
we find that the solutions in the 100-day case are almost the same
despite different bottom boundary conditions. However, the
solutions in the 10-day case are indeed different for different
bottom boundary conditions. Setting the bottom boundary conditions
$p'=T'=0$ at a location well below $\tau_{conv}$ should give rise
to more reasonable solutions for the 10-day case, but would
require a method of treating the dynamics in the convective zone.

The direction of momentum transport is related to the sign of
$\omega$. A retrograde thermal forcing ($\omega<0$) leads to
upward momentum transport. When the sign of $\omega$ is switched
to positive (i.e. the planets spins slowly than the orbit), our
result shows that internal waves transport momentum downwards.

\section{Thermal tides in a rotating planetary atmosphere}

We now consider the linearized dynamics of a thin spherical shell
(a planetary atmosphere) which rotates at the uniform angular
velocity $\Omega$.  We adopt the coordinates $(\theta,\phi,z)$,
where $\theta$ and $\phi$ are the spherical polar angles and $z$
is the altitude. In order to separate the variables and determine
the solutions, we consider the ``perturbations'' to refer to time-
and azimuth-dependent deviations from a spherically symmetrically
irradiated atmosphere. This procedure corresponds to neglecting
the latitudinal dependence of the average irradiation, and
therefore eliminates winds in the basic state. Similar to the
results for the plane-parallel case described in the preceding
section, the damping of the vertically propagating waves should be
able to transport and deposit angular momentum between different
altitudes.

\subsection{Linearized Equations}
We adopt the linearized equations
\begin{equation}
  -\rmi\omega u_\theta'-2\Omega\cos\theta\,u_\phi'=-\f{1}{\rho R_p}\p_\theta p',
\end{equation}
\begin{equation}
  -\rmi\omega u_\phi'+2\Omega\cos\theta\,u_\theta'=
  -\f{\rmi m p'}{\rho R_p\sin\theta},
\end{equation}
\begin{equation}
  0=-\f{1}{\rho}\p_z p'+\f{\rho'}{\rho^2}\p_zp,
\end{equation}
\begin{equation}
  -\rmi\omega\rho'+u_z'\p_z\rho+\rho\Delta=0,
\end{equation}
\begin{equation}
  -\rmi\omega p'+u_z'\p_z p+\gamma p\Delta=-(\gamma-1)\p_zF_z',
\end{equation}
\begin{equation}
  F_z'=F_z\left(\f{\p_zT'}{\p_zT}+\f{3T'}{T}-\f{\rho'}{\rho}-\f{\kappa'}{\kappa} \right),
\end{equation}
\begin{equation}
  \f{p'}{p}=\f{\rho'}{\rho}+\f{T'}{T},
\end{equation}
\begin{equation}
  \Delta=\f{1}{R_p\sin\theta}\p_\theta(u_\theta'\sin\theta)+
  \f{\rmi m u_\phi'}{R_p\sin\theta}+\p_zu_z',
\end{equation}
where all perturbations have the form
\begin{equation}
  \real\left[\bu'(\theta,z)\,\rme^{\rmi m\phi-\rmi\omega t}\right],
\end{equation}
etc.  We have assumed vertical hydrostatic balance and neglected
horizontal radiative diffusion. These assumptions are identical to
those made for the plane-parallel model and are justified in the
Appendix. We have adopted the traditional approximation, which
neglects $2\Omega\sin\theta\,u_z'$ in the  $\phi$-momentum
equation. The traditional approximation is valid if
$|u_z'|\ll|u_\theta'|,|u_\phi'|$ as expected in the atmosphere
where the wave frequency is much smaller than $N$. Under these
assumptions we can solve the horizontal components of the equation
of motion for $u_\theta'$ and $u_\phi'$ and substitute into the
expression for $\Delta$ to obtain
\begin{equation}
  \Delta=-\f{1}{\rmi\omega\rho R_p^2}\mathcal{L}p'+\p_zu_z',
\end{equation}
where $\mathcal{L}$ is the Laplace tidal operator defined by
\begin{equation}
  \mathcal{L}p'=-\f{1}{\sin\theta}(\p_\theta+\nu m\cot\theta)
  \left[\left(\f{\sin\theta}{1-\nu^2\cos^2\theta}\right)
  (\p_\theta-\nu m\cot\theta)p'\right]+\f{m^2p'}{\sin^2\theta},
\end{equation}
with $\nu=2\Omega/\omega$ and in general,
$\omega=m(n_{orb}-\Omega)$. Together with regularity conditions at
the north and south poles, this is a self-adjoint operator with
real eigenvalues $\lambda$ (either positive or negative) depending
on the dimensionless wave frequency $\nu$.  In the non-rotating
limit $|\nu|\ll1$ the eigenfunctions are associated Legendre
polynomials (i.e.~spherical harmonics) $P^m_n(\cos\theta)$ and the
eigenvalues are $n(n+1)$ for integers $n\ge|m|$; more generally
the eigenfunctions are the Hough functions $H_{\nu,m,n}(\theta)$.
It is traditional to express the eigenvalues in terms of an
`equivalent depth' $h$:\footnote{The reason for this name is that
in Laplace's analysis of waves in a shallow incompressible ocean,
the permissible values of $\nu$ for free oscillations are
determined by the condition that $h$ equals the depth of the
ocean.}
\begin{equation}
  \mathcal{L}H_{\nu,m,n}=\lambda_{\nu,m,n}H_{\nu,m,n}=
  \f{\omega^2 R_p^2}{gh_{\nu,m,n}}H_{\nu,m,n}.
\end{equation}

We may then assume that all perturbations other than $u_\theta'$
and $u_\phi'$ (which have been eliminated) depend on $\theta$
through a particular Hough function. This allows for the
separation of variables and we are left with the following system
of ordinary differential equations:
\begin{equation}
  0=-\f{1}{\rho}\p_z p'+\f{\rho'}{\rho^2}\p_zp,
\end{equation}
\begin{equation}
  -\rmi\omega\rho'+u_z'\p_z\rho+\rho\Delta=0,
\end{equation}
\begin{equation}
  -\rmi\omega p'+u_z'\p_z p+\gamma p\Delta=-(\gamma-1)\p_zF_z',
\end{equation}
\begin{equation}
  \Delta=-\f{\lambda}{\rmi\omega R_p^2}\f{p'}{\rho}+\p_zu_z',
\end{equation}
\begin{equation}
  F_z'=F_z\left(\f{\p_zT'}{\p_zT}+\f{3T'}{T}-\f{\rho'}{\rho}-\f{\kappa'}{\kappa} \right),
\end{equation}
\begin{equation}
  \f{p'}{p}=\f{\rho'}{\rho}+\f{T'}{T},
\end{equation}
in which $\mathcal{L}$ has been replaced with the appropriate
eigenvalue $\lambda$.  These are identical to the equations used
for the non-rotating plane-parallel atmosphere and lead to the
same ODEs (eq.~(\ref{l1})-(\ref{l4})) except that the horizontal
wavenumber $k_x$ is replaced by the Hough eigenvalue $\lambda$
according to the formula $k_x^2=\lambda/R_p^2$ (or by the
equivalent depth of the Hough function according to the formula
$k_x^2=\omega^2/gh$). In the non-rotating limit this means
$k_x^2=n(n+1)/R_p^2$, but in the rotating case $k_x^2$ can be
positive (wave solutions) or negative (evanescent solutions).

A WKB analysis of the above linearized equations, in which
$(1/r)\partial_{\theta}$ and $\partial_r$ are replaced by $\rmi
k_{\theta}$ and $\rmi k_r$, gives the dispersion relation for
adiabatic perturbations \citep{OL}
\begin{equation}
k_r^2=\f{N^2}{\omega^2} \left( \f{\lambda}{r^2} \right),
\label{eq:WKB_rot}
\end{equation}
where
\begin{equation}
\f{\lambda}{r^2} \approx \f{k_{\theta}^2}{1-\nu^2 \cos^2 \theta}.
\label{eq:lambda}
\end{equation}
This suggests that the solutions are oscillatory for $\lambda
(1-\nu^2 \cos^2 \theta) > 0$. For solutions with $\lambda >0$, the
oscillations are confined to the equatorial region. This
corresponds to the g-mode solutions modified by rotation
\citep{Longuet,Bildsten}. However, if $\lambda$ is small, this
confinement is relatively weak because $k_\theta$ is imaginary but
small away from the equatorial region. The WKB analysis in the
$\theta$ direction starts to fail in this regime but the radial
(i.e. vertical) wavelength can remain small if $\omega$ is very
small. The solutions in this regime give rise to the baroclinic
Rossby waves, or so called buoyant r modes \citep{Longuet,Heyl}.
In the limit of solutions with very small, or zero, $\lambda$, the
radial WKB approach also fails. These special solutions that are
global in both $r$ and $\theta$ are conventionally referred to as
the barotropic Rossby waves, or simply the Rossby waves or r
modes. Of course, our thin-layer calculation has eliminated
barotropic r modes and we shall see later that non-adiabatic
effects such as thermal diffusion affect the vertical structure of
these modes to some extent.

The above description of the solution properties implies that
$k_x$ is not simply related to the horizontal scale of the tidal
forcing, as in the non-rotating problem. There is a need for
reinterpretation by finding the eigenvalues of the relevant Hough
functions excited by the tidal forcing.
The Hough functions for a given $m$ and $\nu$ can be decomposed in
terms of normalized associated Legendre polynomials $\tilde
P_l^m$. \citep{OL} show that the $n$th eigenvector of a certain
tridiagonal matrix provides the $n$th Hough function in the form
\begin{equation}
H_{\nu,m,n}(\theta) = \sum_l a_{n,l} \tilde P_l^m(\cos(\theta))
\label{eq:Legendre}
\end{equation}
for some coefficients $a_{n,l}$ (the components of the eigenvector
$n$). In the next subsection, we shall demonstrate how we
determine the relevant Hough modes excited by thermal forcing and
describe the associated problem with the method of separation of
variables.

\subsection{Thermal Forcing}
Although the parent star is not distant from its hot Jupiter, we
simplify matters by assuming the stellar irradiation to consist of
parallel light rays impinging on the spherical planet. The stellar
irradiation (heating term) is then proportional to
eq.~(\ref{eq:thermalforcing_plane}) multiplied by $\sin \theta$
with $\tilde\phi=\phi-(n_{orb}-\Omega)t$.

The latitudinal dependence $\sin\theta$ should be decomposed into
Hough functions. If the Hough functions are expressed in a basis
of associated Legendre polynomials with $m=1$, it is necessary to
decompose $\sin\theta$ similarly:
\begin{equation}
\sin \theta = - (2/\sqrt{3}) \tilde P_1^1(\cos\theta) = -
(2/\sqrt{3}) \sum_n b_{1,n} H_{\nu,1,n}(\theta),
\end{equation}
where $b_{1,n}$ is the first row of $b_{l,n}$ which is the inverse
matrix of $a_{n,l}$ in eq.~(\ref{eq:Legendre}). The coefficient
$b_{1,n}$ tells us how much each Hough mode is excited by the
latitude-dependent irradiation with $m=1$ (i.e. the second term on
the right hand side of eq.~(\ref{eq:thermalforcing_plane})).

Having said that the Hough modes are determined from the
latitude-dependent heating in our model, we should note that our
description is not entirely self-consistent because we have
neglected the latitudinal dependence of the average irradiation.
The $\theta$-dependence of $F_i$ would lead to a latitudinal
variation of the properties of the unperturbed atmosphere and
would also generate winds in the basic state. These complications
are in conflict with the method of separation of variables used
for solving the linear problem. However, as we have explained in
the preceding subsection and as we shall see from some examples
later in this paper, the Hough functions in some cases peak at low
latitudes and decay quickly at high latitudes. Moreover, the
latitudinal dependence of $F_i$, $\sin \theta$, is not a fast
varying function of low latitudes. These imply that if we aim to
perform an order-of-magnitude estimate of some wave quantities
integrated over all latitudes (such as the vertical angular
momentum flux computed later in this paper), the contribution of
calculations from high latitudes should be quite small. Therefore,
we may simply apply the unperturbed heating at the equator to all
latitudes and expect that the error introduced from high latitudes
will be diminished by the Hough functions. This allows us to
consider the ``perturbations" referring to time- and
azimuth-dependent deviations from the symmetrically irradiated
atmosphere even though the excited Hough modes are still
determined by the latitude-dependent perturbed heating in our
model.

\subsection{Numerical Results}
We adopt the same input parameters for HD 209458b in the rotating
case as in the plane-parallel case to solve the linear problem. We
consider the diurnal thermal forcing ($m=1$) and the scenario
where the planet rotates faster than its orbit ($\omega <0$).
The Hough functions and eigenvalues are obtained based on
\citet{OL}. We only consider the $n$th ``wave" mode (i.e. $\lambda
>0$)\footnote{However, $\lambda>0$ does not necessarily admit a wave
solution in the vertical direction in our problem involving
thermal diffusion.} which contributes the largest value of
$b_{1,n}$; namely, the largest heating term in the basis of Hough
functions excited by the latitude-dependent diurnal forcing. This
particular $n$ is denoted as $n'$. Then we find that the positive
eigenvalue $\lambda$ associated with the leading Hough mode for
$m=1$ has distinct features between fast and slow thermal tides
(Longuet-Higgins 1968).

\begin{figure}
\epsffile{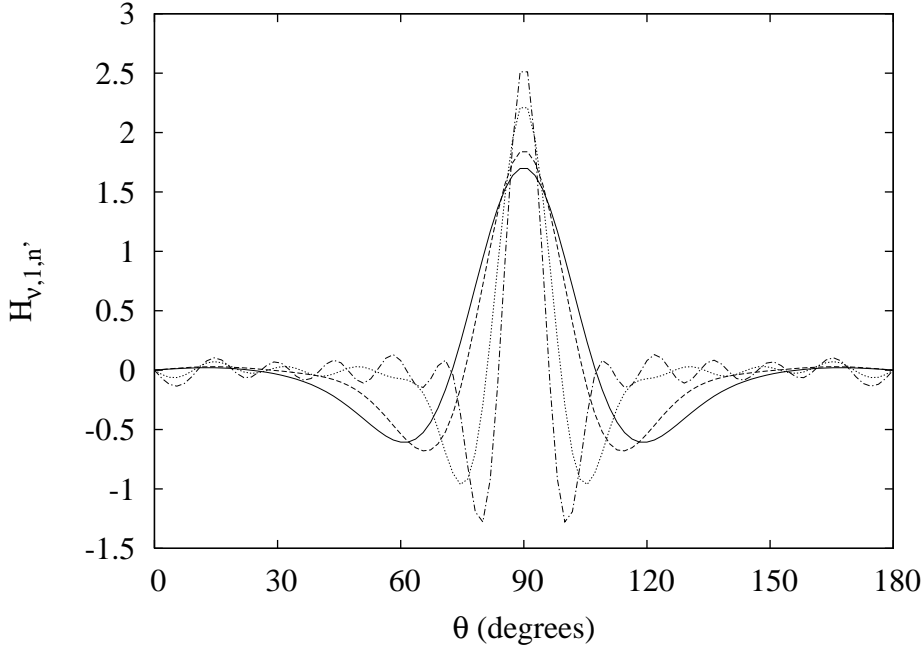} \caption{Hough functions
$H_{\nu,m=1,n=n'}(\theta)=\sum_{l=1}^{l=25} a_{n',l}\tilde
P_l^1(\cos \theta)$ for $-2\pi/\omega= 0.3$ ($\nu \approx -2.17$,
solid curve), $1$ ($\nu\approx -2.57$, dashed curve), $3.5$
($\nu\approx -4$, dotted curve), and $7$ ($\nu=-5.97$, dash-dotted
curve) days.} \label{fig:Hough_fast}
\end{figure}

When $-2\pi/\omega$ is shorter than $\approx 7.04$ days (i.e. $<3$
times the spin period or say $-\nu < 6$), $\lambda$ increases with
the forcing period and is much larger than 1. The dominant Hough
modes in this fast-tide regime have negative $\lambda$ and are
evanescent. Positive and large $\lambda$ are then associated with
less dominant Hough modes which normally consist of more weight
from the associated Legendre polynomials of higher degrees. For
instance, $\lambda$ for the forcing period $=0.3$, 1, 3.5, and 7
days are 37.64, 53.9, 135, and 309 respectively. The corresponding
eigenvectors $H_{\nu,1,n'}$ in the basis of the associated
Legendre polynomials with the normalization coefficient
$\sqrt{[(2l+1)(l-1)!]/[2(l+1)!]}$ summed from $l=1,3,5$ up to 25
are depicted in Figure~\ref{fig:Hough_fast}. These latitudinal
structures agree roughly with the WKB analysis described by
eqs.~(\ref{eq:WKB_rot}) and (\ref{eq:lambda}): as the forcing
period increases, the oscillatory solutions are more equatorially
confined and the latitudinal wavelength becomes shorter. These
waves excited by the fast thermal tides are g modes modified by
rotation.

\begin{figure}
\epsffile{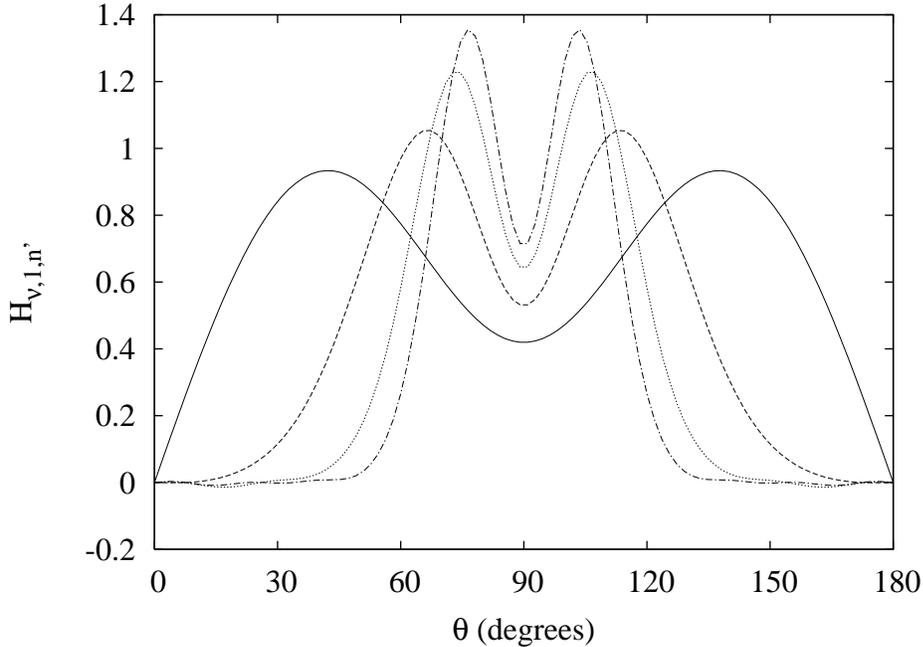} \caption{Hough functions
$H_{\nu,m=1,n=n'}(\theta)=\sum_{l=1}^{l=25} a_{n',l}\tilde
P_l^1(\cos \theta)$ for $-2\pi/\omega= 10$ ($\nu \approx -7.67$,
solid curve), $50$ ($\nu\approx -30.4$, dashed curve), $100$
($\nu\approx -58.7$, dotted curve), and $150$ ($\nu=-87.1$,
dash-dotted curve) days.} \label{fig:Hough_slow}
\end{figure}

When the forcing period is exactly 3 times the spin period,
$\lambda=0$ and the solution of the Laplace tidal equations
without thermal diffusion corresponds to a $m=1$ Rossby wave with
the latitudinal profile $\propto \sin\theta(1+4\cos^2(\theta))$.
When the forcing period $>$ 3 times the spin period, other than
allowing solutions with a very large positive $\lambda$ (i.e.
equatorially confined g modes), the tidal equations also admit
solutions with a small yet positive $\lambda$. They are known as
the barotropic and baroclinic Rossby waves. These new solutions in
the slow-tide regime are the predominant modes excited by the
thermal forcing in our model. The $\lambda$ of the r modes
increases slowly with the forcing period but remains smaller than
1. For instance, the $\lambda$ of the r modes for the forcing
period $=10$, 50, 100, and 150 days are approximately 0.056,
0.109, 0.11, and 0.111 respectively. The Hough functions for these
cases are plotted in Figure~\ref{fig:Hough_slow} to illustrate how
the latitudinal structure of these dominant modes varies with the
forcing period. Note that the r modes are less equatorially
confined than the g modes and therefore couple better with the
global heating profile ($\propto \sin\theta$). This explains why
the r modes are more strongly excited than the g modes in our
model.

\begin{figure}
\epsfxsize=17 cm \epsffile{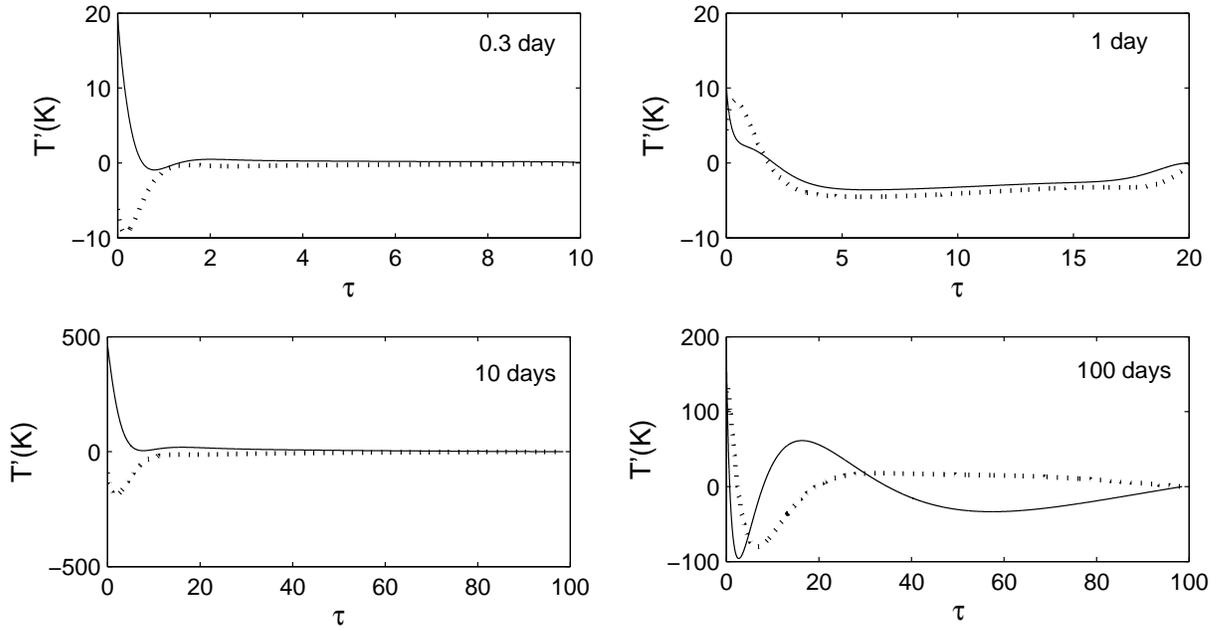} \caption{Same as
Figure \ref{fig:T} except that the Coriolis effect is included.}
\label{fig:T_pert_rot}
\end{figure}

Knowing the eigenvalues $\lambda$ and assuming the unperturbed
irradiation to be spherically symmetrical, we can solve for the
$z$-dependence of the perturbations. For comparison, we start our
study with the rotating cases for the same forcing periods
considered in the non-rotating cases: 0.3, 1, 10, and 100 days.
The results for the temperature perturbations are shown in Figure
\ref{fig:T_pert_rot} (rotating cases) for comparison with Figure
\ref{fig:T} (non-rotating plane-parallel cases). In the 0.3-day
case, the solutions for both the rotating and non-rotating cases
exhibit diffusive behaviour. In the 100-day case, the waves
propagate downwards in the rotating case as in the non-rotating
case but the vertical wavelength in the rotating cases is longer
(therefore the waves are less damped via radiative loss) because
$\lambda$ has a small value. However in the 1-day and 10-day
cases, the Coriolis effect introduces a large and a small
$\lambda$ respectively, turning the diffusive solution to a wave
solution in the 1-day case and turning the wave solution to a
diffusive solution in the 10-day case.

Although the leading modes in the 0.3- and 10-day cases give rise
to vertically diffusive solutions, the less dominant modes (with
therefore smaller magnitudes), do admit vertical wave solutions
due to the much larger values of $\lambda$. These are g modes and
more equatorially confined. For instance, we find that (not
plotted here) the second dominant mode in the 10-day case gives a
vertical wave solution (g mode) because $\lambda \approx 535$ but
its wave magnitude, described by $b_{1,n}$, is 0.024 which is much
smaller than the magnitude $b_{1,n'}=0.905$ of the most dominant
mode. The Hough function in the 10-day case has a latitudinal
profile close to $\sin\theta(1+4\cos^2(\theta))$, more akin to the
barotropic Rossby mode. On the other hand, the dominant modes for
the forcing period $>20$ days do give vertical wave solutions. As
the planet is more rotationally synchronized (i.e., $-\omega$
becomes smaller and therefore $-\nu$ becomes larger), the Rossby
waves are more akin to baroclinic r modes with shorter vertical
wavelength (see the 100-day case in Figure~\ref{fig:T_pert_rot})
and are more equatorially confined. Note that although the r mode
in the 10-day case decays vertically due to thermal diffusion, it
has a broad distribution across latitudes. Therefore the
assumption of symmetrical unperturbed irradiation relying on small
values of Hough functions at high latitudes may not be appropriate
in this case.

\begin{figure}
\epsfxsize=17 cm \epsffile{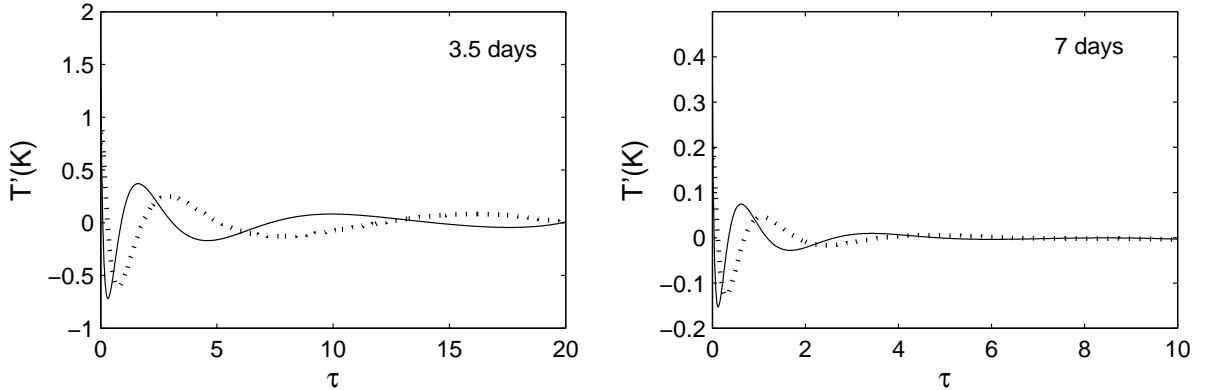} \caption{Same as
Figure \ref{fig:T_pert_rot} but for the rotating cases of
$-2\pi/\omega=3.5$ days and $7$ days.} \label{fig:T_pert_rot2}
\end{figure}

In the regime of fast thermal tides ($-\nu< 6$), only g modes can
exist. In the 1-day case, we have applied the bottom boundary
conditions $p'=T'=0$ at $\tau=20$ to illustrate the results (see
Figure~\ref{fig:T_pert_rot}) because we have difficulty obtaining
the solutions when imposing the bottom boundary conditions at
$\tau=\tau_{conv}$. This may result from the possibility that the
wave of long vertical wavelength in the 1-day case can penetrate
deep into the radiative layer, rendering the bottom boundary
conditions $p'=T'=0$ invalid (cf. the 10-day case in the
non-rotating case). To justify this approach for the 1-day case,
we examine the solutions for forcing periods slightly longer than
1 day, with the expectation that short-wave solutions would appear
owing to the larger $\lambda$. This is illustrated in
Figure~\ref{fig:T_pert_rot2} for the cases of $-2\pi/\omega=3.5$
days ($\lambda\approx 135$) and $7$ days ($\lambda \approx 309$).
As the forcing period increases from 1 day to 3.5 days, and then
to 7 days, the wave solutions can be seen clearly with decreasing
vertical wavelength, as expected.
The larger values of $\lambda$ associated with the less dominant
Hough modes manifest themselves as shorter latitudinal wavelength
as a result of the Coriolis effect, driving internal waves of
shorter vertical wavelength according to the WKB dispersion
relation (see eqs.~(\ref{eq:WKB_rot}) and (\ref{eq:lambda})).

In summary, the diurnal thermal forcing in the rotating case
excites a series of Hough modes. When the forcing period is long
(i.e., $-2\pi/\omega$ is much longer than $\sim 10-20$ days), the
dominant modes are baroclinic Rossby waves that can propagate
downwards. When the forcing period is short (i.e., $-2\pi/\omega$
is shorter than $\sim 10-20$ days), the dominant modes are
evanescent due to either the Coriolis effect (i.e. negative
$\lambda$) or thermal diffusion. However, the less dominant modes
have short latitudinal wavelength and therefore internal waves can
be excited without being subject to the same constraint that the
sound crossing time between the day and night sides needs to be
shorter than the forcing period as in the non-rotating
plane-parallel case.  Since the less dominant modes are small in
magnitude, we expect that the angular momentum transported by the
fast tides (g modes) is smaller than in the case of the slow tides
(r modes). This is the subject of the next subsection.

\subsection{Vertical Angular Momentum Transport}
The vertical angular momentum flux carried by the Hough waves
integrated over the planet's surface is given by \citep{OL}
\begin{equation}
L=\frac{\pi m}{\omega}\int_0^\pi {\rm Re}[{p'}^* u'_z]
R_p^2\sin\theta d\theta,
\end{equation}
where $p'$ and $u'_z$ are both expressed as series of Hough
functions ($p'=\sum_n p'_n H_{\nu,m,n}$, etc.) in our model. Then
the total flux is the sum of the contributions from each Hough
function, because they are orthogonal. Therefore, the angular
momentum flux carried by each Hough mode $H_{\nu,m,n}$ with $m=1$
is given by
\begin{equation}
L_n= \frac{\pi}{\omega}\int_0^\pi {\rm Re}[{p'_n}^*u'_{zn}]
H_{\nu,m=1,n}^2 R_p^2 \sin\theta d\theta =\frac{\pi}{\omega}{\rm
Re}[{p'_n}^*u'_{zn}] R_p^2, \label{eq:ang_mom}
\end{equation}
where we have used the normalization $\int_0^\pi H_{\nu,m,n}^2
\sin\theta d\theta=1$. We focus only on $L_n$ due to the leading
Hough mode (i.e., $n=n'$) and use the peak value of the radial
profile of the angular momentum flux to quantify the torque in
each case of parameter study.

Since our calculation is limited to a thin radiative layer and by
the condition of symmetrical unperturbed irradiation, to
reasonably estimate $L_{n'}$, we focus on the frequency regimes in
which the solutions give short vertical wavelengths and modest
equatorial confinement. Hence we carry out the estimates for the
following forcing periods: 3.5, 7, 100, \& 150 days. The results
are listed in Table~\ref{table}, taking into consideration 4
different cases for each forcing period; Case I: the original case
for the input parameters of HD 209458b, Case II: the same input
parameters as Case I except that a larger $\kappa$ ($c_\kappa$ is
increased by a factor of 100) is used, Case III: the same as Case
I except that a larger $R_p$ ($=2R_J$, a young planet) is used,
and Case IV: the same as Case I except that a smaller $a$ (one
half of $0.0474$ AU) is used. Note that changing $\kappa$, $R_p$,
or $a$ would alter the interior structure and therefore other
input parameters such as $F_z$ need to change accordingly.
However, the purpose of the present case study here is simply to
investigate how $L_{n'}$ varies with each parameter and to better
understand what we can expect from our model for other interesting
cases.

In general, Table~\ref{table} shows that the torques in the
fast-tide cases (i.e. 3.5 and 7 days) are weaker than those in the
slow-tide cases (i.e. 100 and 150 days) by several orders of
magnitude. As described in the previous subsection, this is
primarily because the dominant Hough modes in the fast-tide regime
are evanescent in the vertical direction. More specifically, the
amplitudes of the internal waves $|b_{1,n'}|$ are 0.08 for
$2\pi/\omega=-3.5$ days and 0.04 for $2\pi/\omega=-7$ days, which
are smaller than the amplitudes of the baroclinic Rossby waves
0.88 for $2\pi/\omega=-100$ days and 0.82 for $2\pi/\omega=-150$
days.

Table~\ref{table} also illustrates how different input parameters
affect the torque $L_{n'}$ driven by the thermal tides. Case II
shows that increasing $c_\kappa$ by 2 orders of magnitude only
reduces the torque by less than a factor of 3, indicating that the
result is less sensitive to the opacity. Case III shows that the
torque driven by fast (i.e., 3.5 and 7 days) thermal tides on a
larger hot Jupiter is weaker than that on a smaller hot Jupiter.
The effect is opposite for slow (i.e., 100 and 150 days) thermal
tides. Case IV shows that placing a hot Jupiter closer to its
parent star decreases (increases) the torque in the fast (slow)
tide case. Note that increasing $\kappa$, $R_p$ (therefore $g$),
or $F_i$ reduces the pressure and density of the equilibrium state
at a given $\tau$. This normally leads to an increase in $u'_z$
but does not dictate the change of $p'$ in any unique way among
the various cases. Therefore, the torque does not follow a simple
trend in variation with any of the input parameters.

\begin{table}
\caption{Torques in units of $10^{30}$ dyne cm generated by
thermal tides with $2\pi/\omega=-3.5$, $-100$, and $-150$ days for
4 different cases.} \label{table}
\begin{tabular}{@{}lcccc}
\hline
Case & $-3.5$ days &$-7$ days&$-100$ days & $-150$ days \\
\hline
I (original)      & 0.062 & 0.014 & 3474 & 3011 \\
II (high $\kappa$) & 0.022 & 0.005 & 2276 & 1973 \\
III (large $R_p$) & 0.041 & 0.009& 8414 & 7294 \\
IV (small $a$) & 0.056 & 0.013 & 13719 & 11893 \\
\hline
\end{tabular}
\end{table}

Compared to the torque generated by the gravitational tides, the
torque driven by the radiative damping of the baroclinic Rossby
waves is sizable. With a constant tidal lag angle specified by the
$Q'_p$ value of the planet, the torque due to the gravitational
equilibrium tides driven by a parent star on a non-synchronized
planet is given by (e.g. Goldreich \& Soter 1966)
\begin{eqnarray}
L_{grav}&=&
{3GM_*^2R_p^5 \over 2 a^6 Q'_p} \\
&\approx &2.8\times 10^{32} \left( {M_* \over 1.1 M_{sun}}
\right)^2 \left( {R_p \over 1.32 R_J} \right)^5 \left( {a \over
0.0474 {\rm AU}} \right)^{-6} \left( {Q'_p \over 10^6}
\right)^{-1} \ {\rm dyne\ cm}
\end{eqnarray}
which is comparable to the torques driven by slow Hough waves
listed in Table~\ref{table}. This implies that in our calculation,
thermal tides play a significant role in transferring angular
momentum to the atmospheric gas even if contributions of the
gravitational tides are also a factor in the atmosphere.

\section{Summary and Discussions}
We present a linear perturbation analysis for internal waves
excited by the stellar diurnal thermal forcing ($m=1$ thermal
tides) in a non-synchronized radiative layer of a hot Jupiter. For
computational simplicity, we employ the radiative diffusion
approximation with a power-law opacity throughout our computation
region from the radiative layer to a cloud-free atmosphere and
apply the Marshak boundary condition for energy balance at
$z=+\infty$. We use the parameters of HD 209458b as an
illustrative example.

We first perform a linear perturbation analysis for a non-rotating
plane-parallel atmosphere to explore how the dynamics of thermal
response in the atmosphere varies with the thermal forcing
frequency $\omega$ in the absence of the Coriolis effect. The
boundary conditions at the bottom of the radiative layer are set
to be $p'=T'=0$ which we speculate are less accurate for the waves
with long vertical wavelength. Nevertheless, we find that when the
thermal forcing period is shorter than the sound speed crossing
time between the day and the night sides of the planet, the
periodic thermal heating at the top of the atmosphere diffuses
into a deeper layer with a decay length related to $\omega$. In
the fast thermal forcing regime corresponding to the rotation rate
of our Jupiter, the problem for an atmosphere heated by the
stellar irradiation becomes essentially a 1-D thermal diffusion
problem with $k_x \rightarrow 0$. On the other hand, when the
thermal forcing period is longer than the sound crossing time of
the planet's surface, the day and the night sides are causally
connected and the problem becomes a 2-D phenomenon with
incompressible properties. As a result, the internal waves are
excited at the top of the atmosphere before propagating downwards.
When the planet spins faster (slower) than its orbital motion, the
thermal tides exhibit a retrograde (prograde) motion. The
retrograde (prograde) waves causes the upward (downward) transport
of momentum. The radiative damping of the waves leads to the
deposition of the momentum in the atmosphere.

We then carry out a linear calculation for a thin spherical shell
which rotates at a uniform angular speed. Since $|\omega| \ll N$,
we adopt the traditional approximation and neglect inertial terms
in the vertical momentum equation. We also assume a spherically
symmetrically irradiated atmosphere as the basic state. These
assumptions allow us to separate the variables and obtain the
perturbation of the form Re$[u'(z) H_{\nu,m,n}(\theta)
\exp{(im\phi -i\omega t)}]$. We examine the diurnal thermal
forcing $m=1$ and consider the $n$th component of the Hough
function which contributes the dominant latitudinal structure of
the heating. The linearized ODEs thus remain the same as those in
the plane-parallel case except that the horizontal wavenumber
$k_x$ is replaced by the eigenvalue $\sqrt{\lambda} /R_p$. In
other words, $\lambda$ conceals the information on how the
latitudinal structure is modified by the Coriolis effect. Similar
to the non-rotating plane-parallel case, the internal waves are
driven by thermal forcing at the top of the rotating atmosphere.
However, the internal waves in most of cases are primarily
confined in a band of latitudes close to the equator and therefore
are weakly excited by the global thermal forcing in our model. In
the slow-tide regime (i.e., $-2\pi/\omega$ is much longer than
$\sim 10-20$ days), baroclinic Rossby waves can be largely excited
by the global thermal forcing. When the planet spins faster than
its orbit ($\omega<0$), these waves propagate inwards but
transport angular momentum outwards.
While the torque generated by the radiative damping of the
baroclinic Rossby waves in the slow-tide regime is comparable to
the torque due to the gravitational tides, the torque generated by
internal waves in the fast-tide regime (i.e., $-2\pi/\omega $ is
shorter than $\sim 10-20$ days) is smaller by several orders of
magnitude. The magnitude of the torque is more sensitive to $R_p$
and $a$ than $\kappa$ in our model.

Unlike the thermal tide theories for a dense atmosphere on a
terrestrial planet \citep{Gold, Dobrovolskis, Correia}, our model
for hot Jupiters, which do not have a hard surface, fails to
generate net thermal bulges and therefore causes only internal
transport of angular momentum inside the planet. In fact, the
atmosphere in our model brings the interior closer to, not further
away from, synchronous rotation. Together with the gravitational
tides, one possible equilibrium state for rotation in our scenario
is still the synchronous rotation.

At this point, it is not clear how our linear results without
winds for a non-synchronized hot Jupiter are able to explain the
vertical shear appearing in existing 3-dimensional numerical
simulations for the atmospheres of hot Jupiters. Apparently, the
advection in the upper atmosphere of a hot Jupiter is driven by
the temperature gradient between the day and night sides and
cannot be described by our linear approach. Nevertheless, it is
likely that the wave dynamics plays a role in the lower atmosphere
where the day-night temperature contrast is small. \citet{Cooper}
postulate that the equatorial super-rotating jet occurring in the
deep atmosphere of HD 209458b in their simulation could be due to
wave transport. The numerical simulation by Showman et al. (2008)
for a non-synchronously rotating HD 209458b shows that the
super-rotating jets are more equatorially confined and penetrate
deeper in their case for $2\pi/\omega=-3.5$ days than those in
their cases for $2\pi/\omega=7$ days and $\infty$. Whether the
deeper distribution of the super-rotating jet is due to the Hough
waves driven by retrograde thermal forcing is an interesting
subject worthy of further investigation.

Our work presented in this paper is the first attempt at an
analytical understanding of thermal tides in a hot Jupiter. Our
linear theory suggests that the angular momentum transport in the
atmosphere of a hot Jupiter due to a periodic thermal forcing is
possible and mostly happens in low latitudes through the radiative
damping of Hough waves. This encouraging result lays the
foundation for further improvements in our calculations during
future investigation, such as the inclusion of the semi-diurnal
contribution, the consideration of Kelvin waves (third kind of
Hough waves; e.g. see Longuet-Higgins 1968) driven by prograde
thermal forcing,
or the examination of internal waves driven from clouds in hot
Jupiters. The extension of this work to other applications, for
instance, thermal bulges on hot super-earths or Hough waves on a
pseudo-synchronized hot Jupiter in an eccentric orbit (cf. Langton
\& Laughlin 2008), is necessary for gaining further insight into
the question of how waves can play a role in atmospheric
circulation on hot Jupiters.


\section*{Acknowledgments}

We are grateful to P. Bodenheimer for providing us with the
interior structure data for HD209458b. We thank M.-C. Liang and K.
Menou for useful discussions. We also thank the referee James
Y.-K. Cho for helpful comments to better improve the paper. This
work is supported by the NSC grants in Taiwan through NSC
95-2112-M-001-073MY2 and 97-2112-M-001-017.

\appendix

\section[]{Dimensionless equations and top boundary conditions}
In this section, we demonstrate how the fluid equations are
non-dimensionalized in our problem. This procedure will not only
help to clarify the relative importance among various terms in the
equations more easily, but also shed light on the behaviour of the
solutions near $\tau=0$ and help to specify the top boundary
conditions.

The equilibrium and perturbed states are completely determined
once the parameters $g$, $c_\kappa$, $a$, $b$, $\mu$, $F_z$ and $F_\rmi$ are
specified. Therefore the problem can be non-dimensionalized in
terms of these parameters. We can write $p=\tilde p\, U_p$,
$T=\tilde T\,U_T$, $\rho=\tilde\rho\,U_\rho$, $\kappa= \tilde
\kappa\,U_{\kappa}$, and $z=\tilde
z\,U_z$, where $\tilde p$, $\tilde T$, $\tilde\rho$, $\tilde\kappa$, a
nd $\tilde z$ are dimensionless functions of $\tau$ and the dimensional units
are given by the following relations:
\begin{equation}
  \sigma U_T^4 =F_z,
\end{equation}
\begin{equation}
  U_p=\f{g}{U_{\kappa}},
\end{equation}
\begin{equation}
  U_p=\f{R}{\mu} U_{\rho} U_T,
\end{equation}
\begin{equation}
  U_z=\f{1}{U_{\kappa} U_\rho},
\end{equation}
\begin{equation}
  U_{\kappa}=c_{\kappa} U_p^a U_T^{-4b}.
\end{equation}
The equilibrium solution can then be expressed as
\begin{equation}
  \tilde T=\left[\threequarters(\tau+\twothirds)+f\right]^{1/4}, \end{equation}
\begin{equation}
  {\tilde p}^{a+1}=\f{4}{3} \left( \f{a+1}{b+1} \right)
  \left[ \tilde T^{4(b+1)} -
  \tilde T_\infty^{4(b+1)} \right],
\end{equation}
\begin{equation}
  \tilde\rho=\f{\tilde p}{\tilde T},
\end{equation}
\begin{equation}
  \tilde z=-\int \f{1}{\tilde p^a \tilde T^{-4b}} \f{\rmd\tau}{\tilde\rho},
\end{equation}
where $f=F_\rmi/F_z$ is a dimensionless parameter that determines
the importance of irradiation.  A further dimensionless parameter
associated with the problem is
\begin{equation}
  \epsilon=\f{F_z}{U_p U_c},
\end{equation}
where $U_c\equiv (R U_T/\mu)^{1/2}$, a characteristic velocity
unit. For the parameters adopted for HD 209458b in this paper,
$\epsilon \approx 4.8 \times 10^{-6} \ll1$. We further define an
estimate of the radiative thermal diffusivity $U_\chi$ as follows
\begin{equation}
U_\chi=\f{F_z}{U_\kappa U_\rho U_p}.
\end{equation}

The linearized equations (\ref{l1})-(\ref{l4}) can also be made
dimensionless by writing $\omega=\tilde\omega\,U_\omega$,
$k_x=\tilde k_x\,U_k$, $\xi_z=\tilde\xi_z\,U_z$, $p'=\tilde
p'\,U_p$, $T'=\tilde T'\,U_T$ and $\bF'=\tilde\bF'\,F_z$, where
\begin{equation}
  U_\omega=\f{U_\chi}{U_z^2},
\end{equation}
\begin{equation}
  U_k=\f{U_\omega}{U_c}.
\end{equation}
Note that $U_k U_z = \epsilon$. The above scaling leads to
\begin{equation}
  \p_\tau\tilde\xi_z=(\p_\tau\ln\tilde T-\p_\tau\ln\tilde p)\tilde\xi_z-
  \f{\tilde k_x^2\tilde p'}{\tilde\kappa \tilde\omega^2\tilde\rho^2}+
  \f{1}{\tilde\kappa \tilde\rho}\left(\f{\tilde p'}{\tilde p}-
  \f{\tilde T'}{\tilde T}\right),
\label{eq:ODE1_d}
\end{equation}
\begin{equation}
  \p_\tau\tilde p'=-\epsilon^2 \f{\tilde\omega^2\tilde\xi_z}{\tilde\kappa}+
  \f{1}{\tilde\kappa} \left(\f{\tilde p'}{\tilde p}-\f{\tilde T'}{\tilde T}\right),
\label{eq:ODE2_d}
\end{equation}
\begin{equation}
  \p_\tau\tilde T'=\left[(a+1)\f{\tilde p'}{\tilde p}-(b+1)\f{4\tilde T'}{\tilde T}+
  \tilde F_z'\right]\p_\tau\tilde T,
\label{eq:ODE3_d}
\end{equation}
\begin{eqnarray}
  \p_\tau\tilde F_z'=\tilde p\left[\left(\f{\gamma}{\gamma-1}\right)
  \p_\tau\ln\tilde T-\p_\tau\ln\tilde p\right]\rmi\tilde\omega\tilde\xi_z+
  \epsilon^2\f{\tilde k_x^2\tilde T'}{\tilde\kappa^2 \tilde\rho^2\p_\tau\tilde T} \nonumber \\
  +\f{\rmi\tilde\omega\tilde p}{\tilde\kappa \tilde\rho}\left[\f{\tilde p'}{\tilde
p}-
  \left(\f{\gamma}{\gamma-1}\right)\f{\tilde T'}{\tilde T}\right].
\label{eq:ODE4_d}
\end{eqnarray}
The terms proportional to $\epsilon^2$ may reasonably be omitted.
Neglecting these small terms amounts to assuming vertical
hydrostatic balance and neglecting horizontal radiative diffusion.

For $a+1>0$, we find the behaviour of physically acceptable
solutions as $\tau\to0$ (omitting the $\epsilon^2$ terms) to be as
follows:
\begin{equation}
  \tilde\xi_z=\left[C_1+O(\tau)\right]\ln\tau+\left[C_2+O(\tau)\right],
  \label{eq:tbc1}
\end{equation}
\begin{equation}
  \tilde p'= \tau^{1\over a+1}
  \left[C_3+O(\tau)\right]\ln\tau+\tau^{1\over a+1} \left[C_4 +O(\tau)\right],
\end{equation}
\begin{equation}
  \tilde T'=\left[C_5\tau+O(\tau^2)\right]\ln\tau+\left[C_6+O(\tau)\right],
\end{equation}
\begin{equation}
  \tilde
  F_z'=\tau^{1\over a+1} \left[C_7+O(\tau)\right]\ln\tau+C_8,
\label{eq:tbc4}
\end{equation}
together with the background states
\begin{eqnarray}
\tilde T&=&\tilde T_\infty+\tilde T_1\tau+O(\tau^2), \\
\tilde p^{a+1}&=& {\tilde p_1}\tau +O(\tau^2),
\end{eqnarray}
where $\tilde T_\infty=(\half+f)^{1/4}$, $\tilde
T_1={\textstyle\f{3}{16}}(\half+f)^{-3/4}$, and $\tilde
p_1=(a+1)(\half+f)^b$.
 The only variable to diverge
as $\tau\to0$ is $\tilde\xi_z$, but since $\tilde\rho=\tilde
p/\tilde T=O(\tau^{1\over a+1})$, the mass flux at $\tau=0$
vanishes. In fact the $\epsilon^2$ terms eventually become
important as $\tau\to0$, but we neglect them here.  On
substituting these series into the ODEs, we obtain the following
relations between the coefficients:
\begin{equation}
C_1=(a+1)\left( \tilde T_\infty -\f{\tilde k_x^2 \tilde
T_{\infty}^2}{\tilde\omega^2} \right)\f{1}{\tilde
p_1^{(a+2)/(a+1)}} \left( \f{1}{\tilde T_\infty} \right)^{-4b} C_3
\end{equation}
\begin{equation}
  C_1=-\f{C_2}{a+1}+\left({\tilde T}_{\infty}-\f{\tilde k_x^2 \tilde T_{\infty}^2}{\tilde\omega^2}\right)
  \f{1}{\tilde p_1^{(a+2)/(a+1)}} \left( \f{1}{\tilde T_\infty} \right)^{-4b} C_4
  - \f{1}{\tilde p_1} \left( \f{1}{\tilde T_\infty} \right)^{-4b} C_6,
\end{equation}
\begin{equation}
  C_3=
  -\f{C_6}{\tilde T_\infty}\f{1}{\tilde p_1^{a/(a+1)}} \left( \f{1}{\tilde T_\infty} \right)^{-4b},
\end{equation}
\begin{equation}
  C_5=\f{(a+1)\tilde T_1}{\tilde p_1^{1/(a+1)}}C_3,
\end{equation}
\begin{equation}
  C_7=- \rmi\tilde\omega \tilde p_1^{1/(a+1)}C_1 + (a+1) \rmi\tilde\omega\tilde T_\infty
  \f{1}{\tilde p_1} \left( \f{1}{\tilde T_\infty} \right)^{-4b} C_3.
\end{equation}

In addition, eq.~(\ref{eq:pert_flux}), the thermal forcing
condition at $\tau=0$, in dimensionless terms reads
\begin{equation}
  C_8=8\tilde T_\infty^3C_6-2f', \label{eq:forcing}
\end{equation}
where the thermal forcing term $f'=F_\rmi'/F_z$. As a result, all
the coefficients in the series expansion of the solution can
ultimately be expressed in terms of $C_4$ and $C_6$. We apply the
shooting method in solving the ODEs. We can first guess the values
of $C_4$ and $C_6$ and evaluate the other expansion coefficients
using the above relations.  We then initialize the solution at
$\tau\ll1$ and integrate to $\tau=\tau_{conv}$, where two boundary
conditions can be applied to determine the values of $C_4$ and
$C_6$ and therefore the values of $\tilde \xi$, $\tilde p'$,
$\tilde T'$, and $\tilde F_z'$ at $\tau \ll 1$.

 \label{lastpage}

\end{document}